\documentclass[pdflatex,sn-aps]{sn-jnl}% American Physical Society (APS) Reference Style
\usepackage{amsmath}
\usepackage{amssymb}
\usepackage{amsfonts}
\usepackage{dsfont}
\usepackage{bbm}
\usepackage{physics}
\allowdisplaybreaks

\usepackage{color}
\newcommand{\add}[1]{\textcolor{blue}{#1}}

\newcommand{\cv}[1]{(#1)}%{\left(#1\right)}
\newcommand{\cvb}[1]{[#1]}%{\left[#1\right]}
\newcommand{\cvc}[1]{\{#1\}}%{\left\{#1\right\}}
\newcommand{\cvv}[1]{\vert #1\vert}%{\left\vert #1\right\vert}

\newcommand{\cvr}[1]{\left\langle #1\right\rangle}
\newcommand{\cvdr}[1]{\langle\langle #1\rangle\rangle}
\newcommand{\iden}{\mathbbm{1}}
\newcommand{\prob}{\mathbb{P}}
\newcommand{\oper}[1]{\hat{#1}}

\jyear{2024}%

\begin{document}
\title[Suppression of noise in separation estimation of optical sources]{Suppression of noise in separation estimation of optical sources with spatial-mode demultiplexing}

% \author*[1,2]{\fnm{Fattah} \sur{Sakuldee}\orcidlink{0000-0001-8756-7904}}\email{fattah.sakuldee@sjtu.edu.cn}

% \author*[2,3]{\fnm{{\L}ukasz} \sur{Rudnicki}\orcidlink{0000-0001-8563-6101}}\email{lukasz.rudnicki@ug.edu.pl}

\author*[1,2]{\fnm{Fattah} \sur{Sakuldee}}\email{fattah.sakuldee@sjtu.edu.cn}

\author[2,3]{\fnm{{\L}ukasz} \sur{Rudnicki}}%\email{lukasz.rudnicki@ug.edu.pl}

\affil*[1]{\orgdiv{Wilczek Quantum Center, School of Physics and Astronomy}, \orgname{Shanghai Jiao Tong University}, \orgaddress{\street{\add{800 Dongchuan Road, Minhang}}, \city{Shanghai}, \postcode{\add{200240}}, \country{China}}}

\affil[2]{\orgdiv{The International Centre for Theory of Quantum Technologies}, \orgname{University of Gda\'nsk}, \orgaddress{\street{\add{Jana Ba{\.z}y{\'n}skiego 1A}}, \city{Gda\'nsk}, \postcode{\add{80-309}}, \country{Poland}}}

\affil[3]{\orgdiv{Center for Photonics Sciences}, \orgname{University of Eastern Finland}, \orgaddress{\street{\add{P.O. Box 111}}, \city{Joensuu}, \postcode{\add{FI-80101}}, \country{Finland}}}

\abstract{Spatial mode demultiplexing was proved to be a successful tool for estimation of the separation between incoherent sources, allowing for sensitivity much below the Rayleigh limit. However, with the presence of measurement's noise, superresolution brought by this technique deteriorates rapidly. On a formal ground, this can be seen in terms of, so called, Rayleigh curse known from direct imaging, which while being absent for ideal spatial mode demultiplexing, goes back in a noisy scenario. In this article, we develop a formal procedure to suppress the destructive effect of the noise, proposing a procedure effectively working as an error correction. For noise models given by a random unitary channel generated by a polynomial of creation and annihilation operators, we demonstrate that perfect noise decoupling can be reached by repeating the mode demultiplexers and intervening them by a group of rotations, in the limit of a large number of repetitions and small noise strength. For a special case of displacement noise, our solution is simplified: by using the demultiplexer twice, and interlacing it by a parity operator, given that the noise configuration is frozen between the first and the second step, a perfect decoupling can be achieved. This allows for a recovery of superresolution for a special class of noise generated by displacement operators. Furthermore, for a strong noise correlation between these two steps, our protocol provides an improved measurement resolution.}
		
\maketitle

\section{Introduction}
Quantum theory has been proven to be more advantageous than classical description for information manipulations including metrology, where one can exercise a superresolution measurement. For instance, for the measurement of a two-point separation distance, the superresolution refers to the distinguishability between two points below the Rayleigh limit. Formally speaking, it means that the resolvable separation distance tends to zero (we will often refer to this situation as to the Rayleigh limit itself), while from a more practical point of view we can speak about sub-Rayleigh regime. In recent years, this feature has been extensively theoretically discussed \cite{Tsang2016,Nair2016,Lupo2016,Ang2017,Yu2018,Yang2019,Peng2021,Matlin2022}, and experimentally realized with several optical setups \cite{Paur2016,Tham2017,Zhou2019,Zhang2020,Joshi2022}. 
The superresolution characteristics are enabled by the use of so-called Spatial-Mode Demultiplexing (SPADE), the measurement of spatial mode states, instead of position basis (direct imaging), in which the variance of the SPADE measurement is finite while that of the direct image diverges for vanishing separation (so called Rayleigh's curse). This technique successfully exemplifies quantum advantages in metrology over traditional methods. To our knowledge, most sensitive realization of separation estimation with SPADE reached the distance of two orders of magnitude smaller than the Rayleigh limit, with a sensitivity which is additional three orders of magnitude smaller \cite{Cleme}. 

However, it has also been observed that superresolution may disappear when there is noise in the measurement, i.e. the divergence of the measurement variance in the Rayleigh limit can revive \cite{Gessner2020,Len2020,Oh2021}. 
This singularity occurs even when the noise is weakly perturbing the system, and, although the SPADE measurement gives a better sensitivity than the direct image method for small separation distance $d,$ the divergence is unavoidable. It is then an interesting question of how to modify the SPADE protocol in order to regain the superresolution. 

This brings to our attention a technique of noise suppression called the dynamical decoupling scheme \cite{Viola1998,Viola1999a,Viola1999b}---the insertion of a certain set of operations interlacing a unitary coupled dynamics in order to remove the influence of an external degree of freedom out of the considered system. 
It is proposed that a similar noise canceling protocol can be adapted in the continuous variable systems \cite{Sakuldee2024}. 
In this work, we employ a similar adaptation for the cancellation of the noise in the measurement setup influenced by the imperfection of the mode demultiplexer. The key concept of the protocol is to execute the demultiplexer multiple times and implement a set of manipulations in between two adjacent steps. The intervention operators will disturb the noise operators, and hence allow us to modify it in such a way that the effective operators become null. 
{In this paper we conduct a theoretical analysis of the problem and propose a protocol for generic model of noise, and demonstrate it for the improvement of separation estimation of optical sources with random misalignment. While its exact implementation leading to complete noise cancellation might seem challenging, the presented method shows the directions for experimentally feasible noise reduction techniques.}

The article is organized as follows. The overview of the SPADE measurement is recalled in Sect.~\ref{sec:motive} while the influence of the noise on the measurement is in Sect.~\ref{sec:D-measure}. A noise model is then introduced in Sect.~\ref{sec:measurement-noise} where we also discuss an example of displacement noise in Sect.~\ref{sec:D-noise} which exemplifies the singularity problem. In Sect.~\ref{sec:manipulation}, we introduce a procedure to suppress the noise effect by introducing a set of control operators and repetition of noise channels. We discuss generic mechanisms and their operational descriptions in Secs.~\ref{sec:gen-describe} and in \ref{sec:operation-desc} respectively, and the application of the protocol for the example of displacement noise is given in Sect.~\ref{sec:D-noise-mani}. Finally, conclusions are given in Sect.~\ref{sec:conclusion}.

% %%%%%%%%%%%%%%%%%%%%%%%%%%%%%%%%%%%%%%%%%%%%%%%%%%%%%%
% %%%% PRELIMINARIES
% %%%%%%%%%%%%%%%%%%%%%%%%%%%%%%%%%%%%%%%%%%%%%%%%%%%%%%
\section{Preliminaries}
In this section, we recall the original measurement problem of two points separation distance on the laser beam profile and its sensitivity quantified by Fisher information. An overview of measurement bases, including SPADE, is given.  

\subsection{Measurement Setup}\label{sec:motive}
\begin{figure}[tb]
\begin{center}
\includegraphics[width=0.5\linewidth]{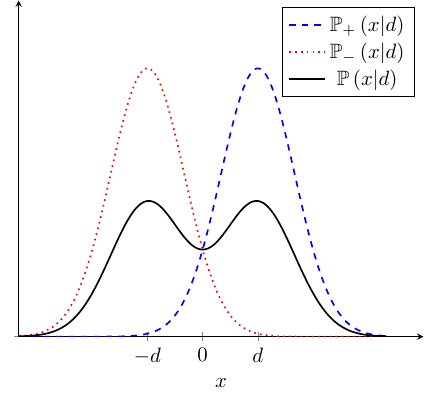}
	\caption{Schematic of Rayleigh's curse scenario introduced in Ref.~\cite{Tsang2016}. The black solid line shows the probability density over position given by the density matrix Eq.~(\ref{eq:rho-init}), i.e. $\prob\cv{x\vert d}:=\Tr\cv{\rho\cv{d}\ketbra{x}{x}}.$ The blue dashed line and red dotted line display the contributions from each term in $\rho,$ i.e. $\prob_\pm\cv{x\vert d}:=\vert\braket{x\vert\pm d}\vert^2.$ In practice this corresponds to the direct measurement of position in which the density probability represents the light intensity at the position $x$ on the image plane. Rayleigh's curse is signified when the distance parameter $d$ approaches the origin, leading to the indistinguishability between the two functions $\prob_\pm\cv{x\vert d}.$}\label{fig:set-up}
\end{center}
\end{figure}

The simplest version of the problem, introduced in Ref.~\cite{Tsang2016}, is to find the value of the displacement $d$ encrypted in a density matrix
	\begin{equation}
		\rho\cv{d} = \frac{1}{2}\cv{\ketbra{+d}{+d} + \ketbra{-d}{-d}} \label{eq:rho-init},
	\end{equation}
where  
    \begin{equation}
        \ket{\pm d}=\displaystyle\int_{-\infty}^\infty dx~\!\psi\cv{x\pm d}\ket{x}, \label{eq:target-states}
    \end{equation}
and
    \begin{equation}
        \psi\cv{x}=\frac{e^{-\frac{{x}^2}{2\sigma^2}}}{\sqrt{2\pi\sigma^2}} \label{eq:target-fns}
    \end{equation}
are Gaussian wave functions centered at $\pm d$ with some variance $\sigma^2$ (see Fig.~\ref{fig:set-up}), and $\ket{x}$ denotes an eigenstate of a position operator $\oper{x}\ket{x}=x\ket{x}.$ 
From now on, we set the variance of the target state $\sigma^2=1,$ in this way fixing units of position and momentum throughout the article. We also write $\rho_{\pm d}=\ketbra{\pm d}{\pm d},$ and hence 
\begin{equation}
    \rho\cv{d}=\frac{\rho_{+ d} + \rho_{- d}}{2}.
\end{equation}

To characterize the parameter $d,$ one can employ measurements given as positive operator valued measures (POVMs) $\mathcal{M}=\cvc{\oper{P}_m},$ $\oper{P}_m\geq 0$ and $\sum_m\oper{P}_m=\iden.$ A construction for the measurement of continuous variables can be done in a similar fashion, while the summation is replaced by integration. Assuming that the measurement can be conducted for arbitrarily many repetitions, the parameter $d$ can be inferred from the statistics over outcomes $\cvc{m},$ i.e. $\prob\cv{m\vert d} = \Tr\cv{\oper{P}_m \rho\cv{d}},$ via standard statistical inference techniques \cite{Rossi2018}. The sensitivity of the inference can be quantified via (classical) Fisher information defined by 
	\begin{equation}
		F\cv{d} = \sum_m \prob\cv{m\vert d}\bigg(\frac{\partial}{\partial d}\ln\prob\cv{m\vert d}\bigg)^2 \label{eq:Fisher-def}
	\end{equation}
which yields a lower bound for a variance $\cv{\Delta d}^2:=\cvr{d^2}-\cvr{d}^2$ through Cram{\'e}r-Rao bound \cite{crameer1946,Rao1992,Rossi2018}
	\begin{equation}
		\cv{\Delta d}^2 \geq F^{-1}\cv{d} \label{eq:Cramer-Rao-bound}.
	\end{equation}
Roughly speaking, the lesser the Fisher information, the more uncertainty in the measurement (inference) of $d.$

Conventionally, the trivial basis of the measurement is given by the projections onto position $\mathcal{M}_x=\cvc{\ketbra{x}{x}},$ where in this case $F\cv{d}\propto d^{2}$ for small $d$. This leads to the problem of Rayleigh's curse or the divergence of the variance Eq.~(\ref{eq:Cramer-Rao-bound}) in the limit $d\rightarrow 0.$  See Fig.~\ref{fig:set-up} for the illustration. In quantum mechanics, this issue can be resolved by choosing other bases, for which the inverse of the Fisher information is not singular. This is the main concept behind the SPADE technique introduced in Ref.~\cite{Tsang2016} where the measurement of two points separation distance is done by using the number (or spatial-mode) basis $\mathcal{M}_{full}=\cvc{\ketbra{n}{n}}_{n=0}^\infty,$ where $\ket{n}$ is an eigenstate of a harmonic oscillator Hamiltonian $\oper{H}=\frac{1}{2}\cv{\oper{x}^2+\oper{p}^2}$ with an eigenvalue $n,$ and $\oper{p}$ denotes a momentum operator. In particular, with these parameter conventions, we write
    \begin{equation}
        \ket{n} = \Big(\frac{1}{2^nn!\sqrt{\pi}}\Big)^{1/2}\int_{-\infty}^\infty dxe^{-x^2/2}H_n\cv{x}\ket{x} \label{eq:state-n},
    \end{equation}
where $H_n\cv{x}$ is a Hermite polynomial of order $n.$ 
With this choice of the basis, it is shown that the Fisher information becomes independent of the parameter $d$ for vanishing separations, being proportional to the classical asymptote $\sigma^{-2}$  \cite{Tsang2016} (or $1$ in our chosen unit). 
The simplest version of this scheme is the binary-SPADE (bi-SPADE), which is the measurement with two coarse-grained states $\mathcal{M}_{1}=\cvc{\oper{R}_0=\ketbra{0}{0},\oper{R}_1=\iden-\oper{R}_0}.$ The Fisher information retains the same value around $d=0$ but decreases to zero asymptotically, and hence it can be adopted for the characterization of $d$ for small values $d<1$, i.e. in the sub-Rayleigh regime. Similarly, the measurement $\mathcal{M}_{K}$ consisting of $K$ lowest-order spatial modes, i.e. $\oper{R}_n=\ketbra{n}{n}$ for $0\leq n<K,$ and the leakage measurement $\oper{R}_K=\iden - \sum_{n=0}^{K-1}\oper{R}_n,$ can be implemented with the same picture.

Although the application of spatial modes as a measurement basis improves the resolution, in practice, it appears that the sensitivity around $d=0$ may collapse when other causes of inaccuracy are introduced to the setup. For instance, this happens when there is noise disturbing the measurement so that the basis of measurement is randomly reshuffled by some stochastic matrix, and hence the measurement statistics become uncertain \cite{Gessner2020}. 
It will be shown later that the effect of such disturbance can limit the advantage of the SPADE technique and pull back the singularity problem for the measurement sensitivity. 
Prior to doing so, let us reformulate the original problem and consider the details of the noise model.

\subsection{Displacement Measurement}\label{sec:D-measure}
First, let us remark that Gaussian wave functions in Eq.~(\ref{eq:rho-init}) can be written as output states of displacement operators generated by a  momentum operator $\oper{D}\cv{\pm d,0} = e^{\mp id\oper{p}}$ acting on a vacuum state $\ket{0}=\displaystyle\int_{-\infty}^\infty dx~\!\psi\cv{x}\ket{x},$ i.e. $\ket{\pm d}=\oper{D}\cv{\pm d,0}\ket{0}.$ In general, we write $\oper{D}\cv{x_0,p_0}:=e^{-ix_0\oper{p}+ip_0\oper{x}},$ where $x_0$ and $p_0$ are real numbers. In this sense, the density matrix Eq.~(\ref{eq:rho-init}) represents the mixture of outputs of two displaced wave functions, in which the parameter $d$ characterizes the separation distance $2d$ between the two. Without loss of generality, we can also define the displacement operator $\oper{D}\cv{z_0}$ with a complex argument $z_0=\tfrac{x_0+ip_0}{\sqrt{2}}$ by writing $\oper{D}\cv{z_0}:=\oper{D}\cv{x_0,p_0}.$ In other words, we introduce standard annihilation $\oper{a}$ and creation operator $\oper{a}^\dagger$ as {$\oper{a}=\tfrac{\oper{x}+i\oper{p}}{\sqrt{2}}$} and then one can write $\oper{D}\cv{z_0}=e^{z_0\oper{a}^\dagger-\overline{z}_0\oper{a}}.$ With this representation it can be said that the displacement acts with respect to the momentum generator (or a translation in position direction) if $z_0$ is real, and to the position generator (or a translation in momentum direction) if $z_0$ is purely imaginary. In the following we will use both: real representation $\cv{x_0,p_0}$ and complex representation $z_0$ interchangeably.

\begin{figure}[!ht]
\begin{center}
\includegraphics[width=0.5\linewidth]{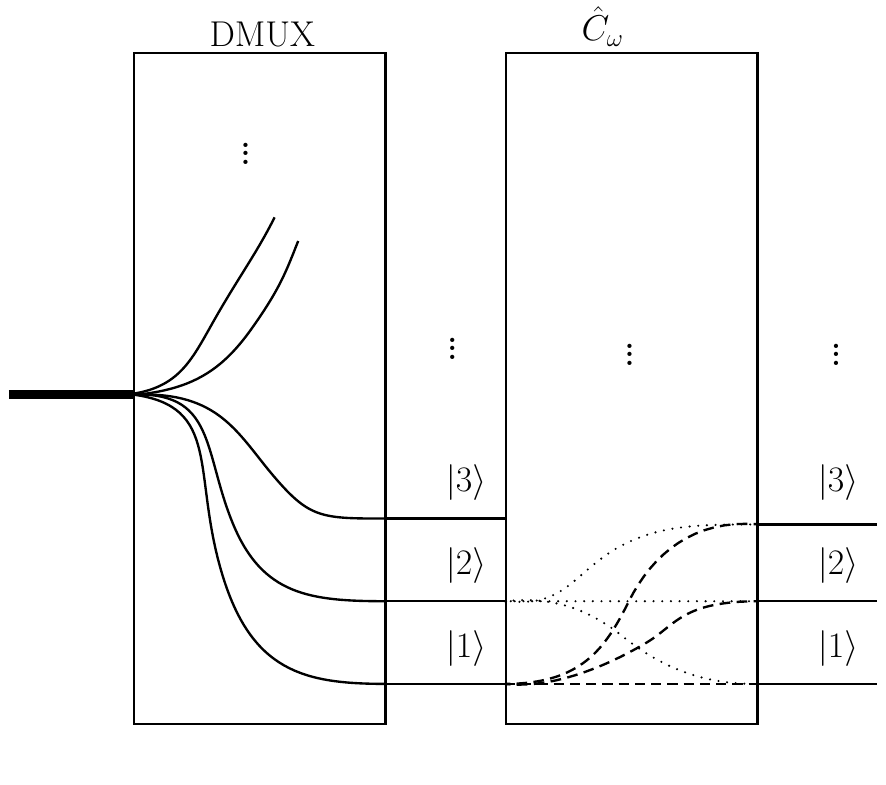}
\caption{Schematic of the spatial demultiplexer and the noise layer model for the noisy channel in the same spirit as in Ref.~\cite{Gessner2020}. In the noiseless case, the input state will pass through the channel and each mode will travel along a different path and exit at the associate port. When the noise is present, the noise layer will interfere with the actual outcomes and introduce crosstalk, leading to misinterpretation of the measurement results.}\label{fig:DMUX-noise-model}
\end{center}
\end{figure}

To model the noise effect, for the SPADE type measurement $\mathcal{M}_K,$ we introduce a random unitary operator $\oper{C}_\omega,$ with some probability space $\cv{\Omega,\mu},$ shuffling the basis $\{\ket{i}:\ket{i}\text{~is an eigenvector of }\oper{R}_i\}$. The subscript $\omega$ denotes a noise configuration  $\omega\in\Omega$. Then
	\begin{equation}
		\prob\cv{n\vert d,\oper{C}_\omega} = \Tr\cv{\oper{R}_n\oper{C}_\omega\rho\cv{d}\oper{C}^\dagger_\omega} \label{eq:P-nd-C}
	\end{equation}
is a probability of noisy measurement for the measurement outcome $n,$ which deviates from the desired profile $\prob\cv{n\vert d}=\prob\cv{n\vert d,\iden}.$ We also write $\prob_\pm\cv{n\vert d,\oper{C}_\omega}=\Tr\cv{\oper{R}_n\oper{C}_\omega\rho_{\pm d}\oper{C}^\dagger_\omega}$.
This noise can naturally appear from the imperfections of the measurement equipment, for instance, the measurement crosstalk or the random shuffling of the measurement outcomes, leading to misinterpretation of the measurement results \cite{Gessner2020}. 
The noise average probability is given by 
    \begin{equation}
		\prob\cv{n\vert d,\oper{C}} := \int_\Omega\prob\cv{n\vert d,\oper{C}_\omega} d\mu\cv{\omega} \label{eq:P-nd-C-ave}.
	\end{equation}
{The integration above is performed over all possible noise configurations $\omega\in\Omega,$ where we sloppily use the symbol $\oper{C}$ for the effects of individual noise operators $\oper{C}_\omega$ and the average over noise configuration. Thanks to linearity, we know that  
    \begin{equation}
        \prob\cv{n \vert d,\oper{C}} = \frac{1}{2}\left(\prob_+\cv{n\vert d,\oper{C}} + \prob_-\cv{n\vert d,\oper{C}}\right). \label{eq:prob_mixture}
    \end{equation}
For the case when the noise can be completely suppressed, $\prob_\pm\cv{n\vert d,\oper{C}}$ is reduced to the same quantity $\prob_+\cv{n\vert d}=\Tr\cv{\oper{R}_n\rho_{+d}}.$
In other words, the noise suppression of the part $\prob_\pm\cv{n\vert d,\oper{C}}$ indicates the noise suppression of the mixture $\prob\cv{m\vert d,\oper{C}}.$ Moreover, it is easy to see that the formulation for $\prob_+\cv{m\vert d,\oper{C}}$ is not technically different from $\prob_-\cv{m\vert d,\oper{C}},$ and hence one can consider the noise suppression procedure for $\prob_+\cv{m\vert d,\oper{C}}$ to determine the noise suppression on the whole mixture.} Therefore, for the rest of this paper we consider the characterization of $d$ from the state $\rho_{+d},$ e.g. via $\prob_+\cv{n\vert d,\oper{C}_\omega}.$

%%%%%%%%%%%%%%%%%%%%%%%%%%%%%%%%%%%%%%%%%%%%%%%%%%%%%%
%%%% MEASUREMENT NOISE AND RELEVANT
%%%%%%%%%%%%%%%%%%%%%%%%%%%%%%%%%%%%%%%%%%%%%%%%%%%%%%
\section{Measurement Noise}\label{sec:measurement-noise}
Here, we elaborate on the model of measurement noise that potentially appears in the setup. In particular, we suppose that the measurement comprises a spatial-mode demultiplexer and an array of photon number resolving detectors, and focus on the noise generated by the first component. Then, we discuss the mathematical structure of the noise in algebraic terms. We illustrate the effect of noise on the measurement sensitivity by a bi-SPADE measurement.

\subsection{Generic Noise Model}\label{sec:noise-model}
In practice, as proposed in the original work Ref.~\cite{Tsang2016}, the measurement operators for SPADE measurements can be described by a spatial-mode demultiplexer, separating incoming beam into different modes, and an array of photon number resolving detectors coupled to the outputs of the demultiplexer and counting photon numbers for each mode. 
The imperfection of both physical components can contribute to an inaccuracy of the measurements. The detector can suffer from photon loss in the apparatus, leading to a disproportion of the signal in each mode. The demultiplexer, on the other hand, may be subject to what so-called crosstalk \cite{Gessner2020}, experienced as misidentification of the photons in output ports, e.g. a photon in mode $\ket{i}$ may appear on the output port associate with the mode $\ket{j},$ creating a false negative detection for the mode $i$ and false positive detection for the mode $j.$

Assuming that the noise distribution is classical, these two types of noise can be treated as parts of the random unitary operator $\oper{C}_\omega,$ where the photon loss contributes to the diagonal terms while the crosstalk affects both diagonal and off-diagonal terms. In this work, we only focus on the crosstalk type of noise since one can execute a self-decoupling principle by repeating the demultiplexing process and inserting an appropriate manipulation operator in between. Note that, by the destructive nature of the photon detectors, the same principle cannot be implemented to correct the noise from the detectors.

Since $\oper{C}_\omega$ is a unitary operator, one can write it in an exponential form as $\oper{C}_\omega=\exp(i\oper{H}_\omega)$ where $\oper{H}_\omega=\oper{H}^\dagger_\omega$ is a Hermitian generator. Now suppose that the noise generator $\oper{H}_\omega$ can be written as a degree $m$ polynomial in the creation  and annihilation operators, i.e.,
    \begin{equation}
		\oper{H}_\omega=\oper{H}_\omega\cv{b_m,\ldots,b_1}= \sum_{k=1}^m\big[b_k\cv{\omega}{\oper{a}}^k+{b}^*_k\cv{\omega}\cv{\oper{a}^\dagger}^k\big] \label{eq:element-G-non-Gauss},
    \end{equation}
where coefficients $b_k\cv{\omega}$ and their conjugates ${b}^*_k\cv{\omega}$ are random complex numbers associated with the configuration $\omega.$ The expression above covers several interesting physical disturbances. For instance, for $m=1$ it is a displacement noise, describing a random shift in the phase space, and for $m=2$ with $b_1=0,$ the noise operator describes the squeezing noise inducing random changes in the variance of the incoming beam \cite{Sakuldee2024}.

{Although, the form of noise generators in Eq.~(\ref{eq:element-G-non-Gauss}) does not include cross-terms, it will be seen later in Sec~\ref{sec:gen-describe} that the noise generators with and without cross-terms can be controlled by the same decoupling group, and the expected decoupling protocols will be the same for both cases.} Hence it suffices to consider the simplified form and the general case with cross-terms can be straightforwardly derived. In this work, we focus on Eq.~(\ref{eq:element-G-non-Gauss}) to illustrate our manipulation protocol. We will revisit this again after discussing the effect of noise on the sensitivity in the next example.

\subsection{Example: Displacement Noise in bi-SPADE Protocol}\label{sec:D-noise}
Here we demonstrate a singularity problem with a displacement noise equipped with Gaussian distribution. It can be seen that, even with tiny noise strength, the singularity appears. 
First, recall the probability for the noisy SPADE measurement $\mathcal{M}_{full}:$
	\begin{equation}
		\prob_+\cv{n\vert d,\oper{C}_\omega} = \Tr\cv{\oper{R}_n\oper{C}_\omega\rho_{+d}\oper{C}^\dagger_\omega} \label{eq:D-measure-prob}.
	\end{equation}
As previously discussed, the most trivial class of the unitary operator $\oper{C}_\omega$ is the class of displacement operators. 
An average displacement noise map on a state $\rho$ reads
	\begin{equation}
		\mathcal{C}_D\cvb{\rho} = \int_\Omega \oper{D}\cv{z_\omega}\rho\oper{D}^\dagger\cv{z_\omega} d\mu\cv{\omega} \label{eq:D-noise-gen},
	\end{equation}
where $z_\omega$ is a random complex number equipped with probability space $\cv{\Omega,\mu}.$ Despite its triviality, it is the most intuitive type of noise for displacement measurement problems, since it generates uncertainty by shifting the center of the target state randomly before subjecting it to the measurements. 

For illustration, the whole process can be considered as the following: first, the state of the system is prepared in a vacuum state $\ket{0};$ then the target quantity $d$ is encoded to the state via the displacement operator in the position direction $\oper{D}\cv{d};$ after this encoding procedure, the center of the out-coming profile will be shifted further by $\oper{D}\cv{ z_\omega}$ equipped with the distribution $\mu;$ finally the state is projected onto the mode $\ket{n}$ and the probability Eq.~(\ref{eq:D-measure-prob}) is calculated to infer the displacement parameter $d.$ 

For generic distribution, one can observe that
	\begin{align}
		\prob_+\cv{n\vert d,\oper{C}} &=\int_{\Omega} \big\vert\bra{n}\oper{D}\cv{d+ z_\omega}\ket{0}\big\vert^2d\mu\cv{\omega}\nonumber\\
		    &= \frac{1}{n!} \int_{\Omega} e^{-\cvv{d +  z_\omega}^2}\cvv{d+ z_\omega}^{2n} d\mu\cv{\omega}\nonumber\\ 
		    &= \Bigg[\frac{\partial^n}{\partial^n a}\frac{\cv{-1}^n}{n!}\int_{\Omega} e^{-a\cvv{d+ z_\omega}^2} d\mu\cv{\omega}\Bigg]_{a=1}\label{eq:integration-compact-form}.
	\end{align}
For the first equality, we employ an identity of displacement operators $\oper{D}\cv{z_1}\oper{D}\cv{z_2}=e^{\cv{z_1\overline{z}_2-\overline{z}_1z_2}/2}\oper{D}\cv{z_1+z_2}$ and the phase term vanishes within the absolute value since one of the argument, $d,$ is a real number. The second equation is derived from the inner product of the Gaussian function of variance $1$ centered at $d+ z_\omega$ with the state $\ket{n}$ given in Eq.~(\ref{eq:state-n}). Now suppose that configuration space $\omega$ per se is the phase space $\mathbb{R}^2$ containing $\cv{x_\omega,p_\omega}$ corresponding to $z_\omega,$ and the disturbance profile is given by a Gaussian distribution function
    \begin{equation}
        d\mu\cv{\omega} = \frac{1}{2\pi\tilde{\sigma}^2}\exp\Big[-\frac{x_\omega^2+p_\omega^2}{2\tilde{\sigma}^2}\Big]dx_\omega dp_\omega \label{eq:noise-gaussian-profile}.
    \end{equation}
The probability of finding the outcome $n$ for the measurement $\mathcal{M}_K$ reads
\begin{align}
    \prob_+\cv{n\vert d,\oper{C}} &=\frac{\cv{-1}^n}{2n!\pi\tilde{\sigma}^2}\Bigg[\frac{\partial^n}{\partial^n a}\int_{-\infty}^\infty\int_{-\infty}^\infty dx_\omega dp_\omega e^{-a\cv{d+ x_\omega}^2 -ap_\omega^2-\frac{x_\omega^2+p_\omega^2}{2\tilde{\sigma}^2}}\Bigg]_{a=1}\nonumber\\
    &=\Bigg[\frac{\partial^n}{\partial^n a}\frac{\cv{-1}^n\exp\cv{-\frac{ad^2}{1+2a\tilde{\sigma}^2}}}{n!\cv{1+2a\tilde{\sigma}^2}} \Bigg]_{a=1}  \label{eq:prob-Gauss-shift}.
\end{align}
	
\begin{figure}[!ht]
\begin{center}
	\includegraphics[width=0.7\linewidth]{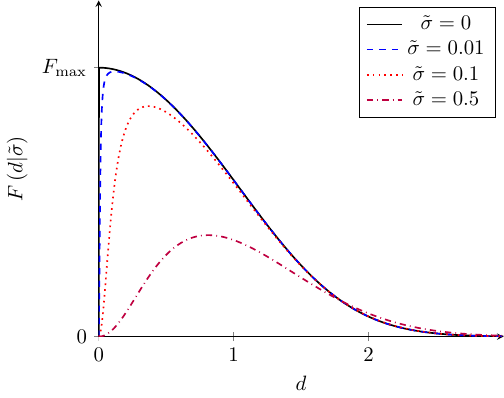}
	\caption{Fisher information in bi-SPADE protocol under Gaussian distributed displacement noise with different noise strengths $\tilde{\sigma}=0, 0.01, 0.1$ and $0.5.$ We term $F_{\max}=F_0\cv{0},$ the maximal Fisher information possible in the bi-SPADE procedure (and noiseless SPADE procedure.)}\label{fig:F-d-Gauss}
\end{center}
\end{figure}

For bi-SPADE procedure we arrive at Bernoulli probability $p_0 = \frac{e^{-d^2/\cv{1+\tilde{\sigma}^2}}}{{1+\tilde{\sigma}^2}}$
and $p_1=1-p_0.$ Taking Fisher information defined in Eq.~(\ref{eq:Fisher-def}), we get
	\begin{equation}
		F\cv{d\vert \tilde{\sigma}} = \frac{4d^2g_{\tilde{\sigma}}^3e^{-g_{\tilde{\sigma}}d^2}}{1-g_{\tilde{\sigma}}e^{-g_{\tilde{\sigma}}d^2}} \label{eq:F-shift-P-Gauss},
	\end{equation}
where $g_{\tilde{\sigma}}=\frac{1}{1+\tilde{\sigma}^2}.$ For the weak noise $\tilde{\sigma}\ll 1,$ given that the distance $d$ is not minuscule, we have
	\begin{equation}
		F\cv{d\vert \tilde{\sigma}} \approx F_0\cv{d}+\frac{  d^2-3 +2e^{-d^2}}{1-e^{-d^2}}F_0\cv{d}\tilde{\sigma}^2 \label{eq:F-shift-P-Gauss-approx},
	\end{equation}
and the first term $F_0\cv{d}=\frac{4 d^2e^{-d^2}}{1-e^{-d^2}}$ represents the Fisher information in bi-SPADE noiseless procedure. See Fig.~\ref{fig:F-d-Gauss} for illustration.

Here an interesting phenomenon appears, namely the interplay between $\tilde{\sigma}$ and $d,$ i.e. the prefactor of $\tilde{\sigma}^2$ diverges in the limit $d\rightarrow 0,$ leading to the issue of scaling. The order of taking two limits, $d\rightarrow 0$ and $\tilde{\sigma}\rightarrow 0,$ is not interchangeable in this scenario, that is, from Eq.~(\ref{eq:F-shift-P-Gauss}), we see that
\begin{equation}
    \displaystyle\lim_{d\rightarrow 0}\lim_{\tilde{\sigma}\rightarrow 0}F\cv{d\vert \tilde{\sigma}}\neq 0\qquad \mathrm{but}\qquad\displaystyle\lim_{\tilde{\sigma}\rightarrow 0}\lim_{d\rightarrow 0}F\cv{d\vert \tilde{\sigma}}= 0. 
\end{equation}
Furthermore, the expansion of $F\cv{d\vert \tilde{\sigma}}$ as a polynomial of $d$ after the expansion in $\tilde{\sigma}$ (or $\tilde{\sigma}$ after $d$) is not the Taylor series of two variables (a qualitatively the same conclusion was reached in \cite{Linowski2023}).

One of the physical examples concerning displacement noise discussed above is the random misalignment in the optical setup of the measurement. The problem is also discussed in the original work Ref.~\cite{Tsang2016}, where the adaptive continuous measurement conditioned on the measurement history is used to avoid the noise and recover the superresolution. A similar concept is considered in Ref.~\cite{deAlmeida2021} in which the correction operator is optimized using the statistical inference technique. In this work, we consider a different perspective of the problem and propose a self-correction protocol by modifying the measurement setup.

From this example, which signifies the collapse of Fisher information at the origin, one could expect a similar conclusion for other types of distribution or even the different models to share this issue and then the superresolution may not be achieved with the presence of noise disturbance. In the following, we will introduce the procedure for noise suppression for this specific class concerning displacement noise.

%%%%%%%%%%%%%%%%%%%%%%%%%%%%%%%%%%%%%%%%%%%%%%%%%%%%%%
%%%% MEASUREMENT MANIPULATION AND NOISE CANCELLATION
%%%%%%%%%%%%%%%%%%%%%%%%%%%%%%%%%%%%%%%%%%%%%%%%%%%%%%
\section{Measurements Manipulation and Noise Cancellation}\label{sec:manipulation}
In this section, we propose a procedure for noise suppression in SPADE measurement. We briefly overview the group-theoretic description of the decoupling mechanism and demonstrate that a set of rotations can be considered a control group for our noise model. We discuss the operational picture of the controls and propose a protocol to execute the noise decoupling operator in our problem. Lastly, we revisit the example of displacement noise in Sect.~\ref{sec:D-noise} and demonstrate how our protocol can provide a perfect decoupling or a better resolution in the small parameter regime.

\subsection{General Description}\label{sec:gen-describe}
First, we discuss the formulation for noise cancellation in Ref.~\cite{Sakuldee2024}. Recall Eq.~(\ref{eq:element-G-non-Gauss}) for the generator of $\oper{C}_\omega$
    \begin{equation}
		\oper{H}_\omega=\oper{H}_\omega\cv{b_m,\ldots,b_1}= \sum_{k=1}^m\big[b_k\cv{\omega}{\oper{a}}^k+{b}^*_k\cv{\omega}\cv{\oper{a}^\dagger}^k\big].\nonumber
    \end{equation}
It is an element of a noise space $\mathcal{I}_S$ generated by $\cvc{\oper{a},\oper{a}^\dagger,\cv{\oper{a}}^2,\cv{\oper{a}^\dagger}^2,\ldots,\cv{\oper{a}}^m,\cv{\oper{a}^\dagger}^m}.$ Note that all the noise operators $\oper{C}_\omega$ belong to this space regardless of their random feature. Now, we construct a control group  
    \begin{equation}
        \mathcal{G}_m=\cvc{(\oper{g}_1)^j=\oper{R}_{\frac{j\pi}{m}}}_{j=0}^{2m-1} \label{eq:G-group-non-Gaussian},
    \end{equation}
generated by $\oper{g}_1=\oper{R}_{\frac{\pi}{m}}=e^{-i\pi\oper{a}^\dagger\oper{a}/m}$ and we denote the identity by $\oper{g}_0=\iden.$ From the formulation suggested in Ref.~\cite{Sakuldee2024}, it is known that
    \begin{equation}
		\sum_{j=0}^{2m-1}\cv{\oper{g}^\dagger_1}^j{\oper{a}}^p\oper{g}_1^j  
		= \Bigg(\frac{1-e^{2ip\pi}}{1-e^{ip\pi/m}}\Bigg)\oper{a}^p=0 \label{eq:ap-erase}
    \end{equation}
for $p\neq 2m,$ or in other words,
    \begin{equation}
		\frac{1}{\cvv{\mathcal{G}_m}}\sum_{\oper{g}_j\in\mathcal{G}}\oper{g}^\dagger_j\mathcal{I}_S\oper{g}_j=0 \label{eq:non-Gaussian-G-I_S-ave},
    \end{equation}
where $\cvv{\mathcal{G}_m}=2m$ denotes a carnality of the group $\mathcal{G}_m.$ 

{When there appear cross-terms in Eq.~(\ref{eq:element-G-non-Gauss}) we can see that the mechanism as in Eq.~(\ref{eq:ap-erase}) also holds in all cases, except when the exponents of creation and anihilation operators are equal, i.e. the terms that are the powers of the operator $\oper{a}\oper{a}^\dagger.$ In other words, the average of a cross-term $\oper{a}^p\cv{\oper{a}^\dagger}^q,$ which can be obtained by replacing $\oper{a}^p$ with such a cross-term in Eq.~(\ref{eq:ap-erase}), vanishes as well whenever $p-q\neq2m$ and  $p-q\neq 0.$ The former case is forbidden by construction, whereas, the latter case leads to a trivial shift in the global phase. In particular,
    \begin{equation}
		\sum_{j=0}^{2m-1}\cv{\oper{g}^\dagger_1}^j{\oper{a}}^p \cv{\oper{a}^\dagger}^{q}\oper{g}_1^j  = \left\{\begin{array}{lr}
		      0,  & p\neq q,  \\
		      2m\cv{\oper{a}\oper{a}^\dagger}^p, & p = q.
		\end{array}\right.
		   \label{eq:ap-erase-cros-term}
    \end{equation}
For the first case, $p\neq q,$ the decoupling can be done as in our main Hamiltonian while the remaining terms with $\cv{\oper{a}\oper{a}^\dagger}^p$ for any $p$ cannot be seen in the measurement since they commute with the measurement operators $\hat{R}_n,$ i.e., $e^{i\varphi\cv{\oper{a}\oper{a}^\dagger}^p}\hat{R}_ne^{-i\varphi\cv{\oper{a}\oper{a}^\dagger}^p}=\hat{R}_n$ for any real number $\varphi.$} 
In this sense, although we consider the decoupling protocol for the generators without cross-terms, in principle, the results can also be applied to the more generic polynomial forms without difficulty.

The property Eq.~(\ref{eq:non-Gaussian-G-I_S-ave}) is the main ingredient for self-cancellation in our protocol.
To see that, let us consider a simple scenario of a $2m$ times concatenation of the operator $\oper{C}_\omega$ with possibly different configuration $\omega,$ i.e., $\prod_{j=0}^{2m-1}\oper{C}_{\omega_j}= \prod_{j=0}^{2m-1}e^{i\oper{H}_{\omega_j}}$ where $\omega_j$ is a noise configuration for the step $j.$ Clearly, the noise strength or the norm of the generator is increasing, and so is the noise variance on the target measurement. However, if we insert the elements of the control group $\mathcal{G}_m$ interlacing the product in such a way that 
    \begin{equation}
        \prod_{j=0}^{2m-1}\oper{C}_{\omega_j} \mapsto \prod_{j=0}^{2m-1}\cv{\oper{g}^\dagger_j\oper{C}_{\omega_j}\oper{g}_j} \label{eq:maps-C-naive},
    \end{equation}
{the effective generator can be modified and approximatly eradicated, as discussed in the following.}

First, let us introduce a scaling parameter $\lambda$ in the noise parameters $b_k(\omega)\mapsto\lambda b_k(\omega),$ or in other words $\oper{H}_{\omega_j}\mapsto\lambda\oper{H}_{\omega_j}.$ For identical configuration $\omega_j=\omega$ for all $j,$ one can directly write the product above in a small $\lambda$ regime
    \begin{equation}
        \prod_{j=0}^{2m-1}\cv{\oper{g}^\dagger_j\oper{C}_\omega\oper{g}_j} = \exp\Bigg[i\lambda\sum_{j=0}^{2m-1}\oper{g}^\dagger_j\oper{H}_{\omega}\oper{g}_j\Bigg] +\mathcal{O}\cv{\lambda^2}\label{eq:C-small-lambda}.
    \end{equation}
Since $\oper{H}_{\omega_j}$ is an element of the noise space $\mathcal{I}_S,$ Eq.~(\ref{eq:non-Gaussian-G-I_S-ave}) simply implies that 
    \begin{equation}
        \prod_{j=0}^{2m-1}\cv{\oper{g}^\dagger_j\oper{C}_\omega\oper{g}_j} = \iden+\mathcal{O}\cv{\lambda^2}\longrightarrow \iden  \label{eq:prod-C-lambda-limit},
    \end{equation}
as $\lambda\rightarrow 0.$ {We call the application of decoupling group Eqs.~(\ref{eq:maps-C-naive})-(\ref{eq:C-small-lambda}) as one cycle of noise decoupling. For a more general case $\omega_j\neq \omega_l$ for $j\neq l,$ one can employ multiple cycles of the procedure above for noise suppression.} For instance, assuming that the noise statistic is ergodic \cite{Sakuldee2024}, by repeating  {the protocol Eqs.~\ref{eq:maps-C-naive}-\ref{eq:C-small-lambda} for $N$ cycles,} one can achieve the product
    \begin{align}
        \prod_{l=1}^{N}\Bigg[\prod_{j=0}^{2m-1}&\cv{\oper{g}^\dagger_{j\oplus l}\oper{C}_{\omega_{j+l}}\oper{g}_{j\oplus l}}\Bigg] = \exp\Bigg(iN\lambda\sum_{j=0}^{2m-1}\oper{g}^\dagger_j\overline{H}\oper{g}_j\Bigg)+\mathcal{O}\cv{\lambda^2} \label{eq:C-small-lambda-N},
    \end{align}
where $j\oplus l:= \cv{j+l} \mod 2m,$ and $\overline{H}=\frac{1}{N}\sum_{l=1}^{N}\oper{H}_{\omega_{j+l}}$ is an ergodic average generator (which is independent of the label $j.$) With a similar argument, Eq.~(\ref{eq:C-small-lambda-N}) will approach the identity operator as in the identical case \cite{Viola1999a,Sakuldee2024}. In other words, with appropriate scaling and conditions, one can induce a noise self-cancellation by inserting a sequence of control operators constructed from the structure of the noise operator.

Now let us briefly discuss the scaling parameter $\lambda.$ In principle, this can be related to the random strength of the noise in the setup. For instance, in Ref.~\cite{Gessner2020}, it is introduced that the noise strength plays an important role in the resolution profile of the measurement. The effects of the noise in terms of this parameter $\lambda$ in this spirit is extensively studied in Refs.~\cite{Linowski2023,Schlichtholz2024}. 
In sub-Rayleigh's regime, when the parameter $d$ is also small, the interplay between $d$ and the noise strength is more important. For the example in Sect.~\ref{sec:D-noise} the coupling constant $\lambda$ can be determined from the noise variance $\tilde{\sigma}^2,$ i.e., $\lambda=\tilde{\sigma}^2,$ one can see that $\displaystyle\lim_{d\rightarrow 0}\lim_{\tilde{\sigma}\rightarrow 0}F\cv{d\vert \tilde{\sigma}}\neq\lim_{\tilde{\sigma}\rightarrow 0}\lim_{d\rightarrow 0}F\cv{d\vert \tilde{\sigma}}.$ This is due to the non-homogeneity of the problem. We stress here that, as we mean by the limit of small noise strength $\lambda,$ the limit $\lambda\rightarrow 0$ should be always taken before the limit of the parameter $d\rightarrow 0.$

\subsection{Operational Description of the Manipulation Protocol}\label{sec:operation-desc}
Here we will translate the noise cancellation mechanism above into the setup of our problem. The main object of the scheme is the control generator $\oper{g}_1=e^{-i\pi\oper{a}^\dagger\oper{a}/m}.$ First, we observe that the number states are its eigenvectors, and then one can write $\oper{g}_1=\sum_{n=0}^\infty e^{-i\pi n/m}\ketbra{n}{n}.$ Here, one faces a degeneracy, namely, $\oper{g}_1\ket{n}=\oper{g}_1\ket{n+2mk}$ for any integer $k.$ In other words, one can write $\oper{g}_1=\sum_{n=0}^{2m-1}e^{-i\pi n/m}\oper{P}_n,$ where $\oper{P}_n=\sum_{k\in \cvb{n}_{2m}}\ketbra{k}{k}$ and $\cvb{n}_{2m}:=\{k:k-n\text{~is divisible by~}2m\}.$ 
The other operators $\oper{g}_j,$ as well as their product, e.g. $\oper{g}_{j+1}\oper{g}_j,$ are also constructed in a similar way with different groupings and phase factors.

\begin{figure}[!ht]
\centering
\includegraphics[width=0.5\linewidth]{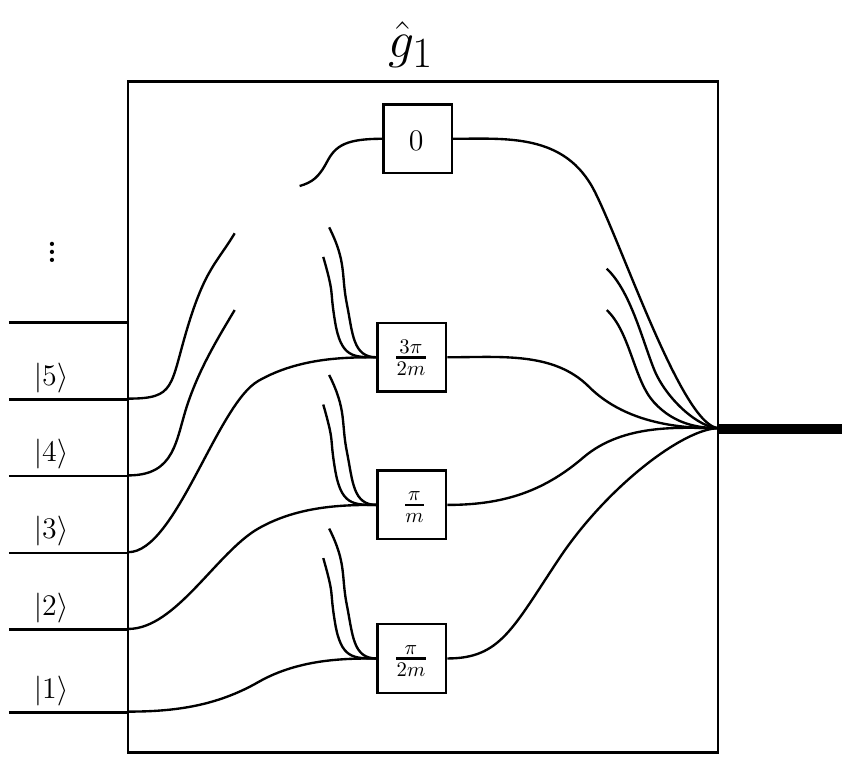}
\caption{Schematic of the primitive control operator $\oper{g}_1$ in terms of mode modulators. This can be realized by grouping all the modes within the same degeneracy subspace with respect to $\oper{g}_1$ and shifting their phases collectively by the eigenvalue argument corresponding to such subspace. For instance, modes $\ket{1},\ket{2m+1},\ket{4m+1},\ldots$ will be assigned with the phase $\tfrac{\pi}{2m},$ modes $\ket{2},\ket{2m+2},\ket{4m+2},\ldots$ will be assigned with the phase $\tfrac{\pi}{m},$ and modes $\ket{2m},\ket{4m},\ket{6m},\ldots$ will be assigned with the trivial phase $0.$ All the control operators $\oper{g}_j$ and their products can also be interpreted in a similar fashion.}\label{fig:g1}
\end{figure}

With this at hand, one can realize the operator $\oper{g}_1$ (as well as other $\oper{g}_j$) by at most $2m$ gates. In fact, since for each degeneracy subspace, the operator acts as a multiplicative operator by a phase factor, it can be prepared by simply grouping the outputs and passing them to corresponding phase shifters. As depicted in Fig.~\ref{fig:g1}, the mechanism begins with separating the incoming modes in equivalence classes (modulo $m$) mentioned earlier, then follows by superposing the beams of the same class and multiplying with the phase determined from its eigenvalue for each class. For instances, for the class $\cvb{1}$ or the modes $\ket{1},\ket{2m+1},\ket{4m+1},\ldots$ their phases will be shifted by $\tfrac{\pi}{2m},$ and so on. After the multiplication, all beams are grouped and pass through the demultiplexer in the next step and are subject to the noise in the next round of repetition in the protocol. We mention here that all other control operators $\oper{g}_j$ can also be prepared in the same way with different rules of superposition and phase shifting.

\begin{figure}[!ht]
\centering
\includegraphics[width=0.4\linewidth]{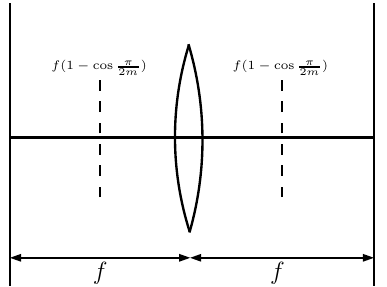}
\caption{A practical implementation of the primitive control operator $\oper{g}_1.$ The left plane, represented by the dashed line, is the input plane while the output plane resides at a distance $f(1-\cos\tfrac{\pi}{2m})$ on the other side from the converging lens with focal length $f.$}\label{fig:g1_lens}
\end{figure}

In practical terms, the operator $\oper{g}_1$ above can also be implemented by a lens system. In fact the $\oper{g}_1$ is simply a rotation or a fractional Fourier transformation operator \cite{Stoler1981,Ozaktas1995,Jagoszewski1998}. As shown in Fig.~\ref{fig:g1_lens}, one can place a converging lens to perform an action of the operator $\oper{g}_1$ where the profile at the out-of-focus planes will correspond to the input and output of the rotation $\oper{g}_1.$ For example, in Ref.~\cite{Jagoszewski1998}, it is suggested that the other control operators $\oper{g}_j$ can also be prepared with the same technique, where the corresponding distance is $f(1-\cos\tfrac{j\pi}{2m})$ for $1\leq j\leq 2m$ (it can be larger than the focal length for $j>m.$) 

\begin{figure}[!ht]
\begin{center}
\includegraphics[width=0.65\linewidth]{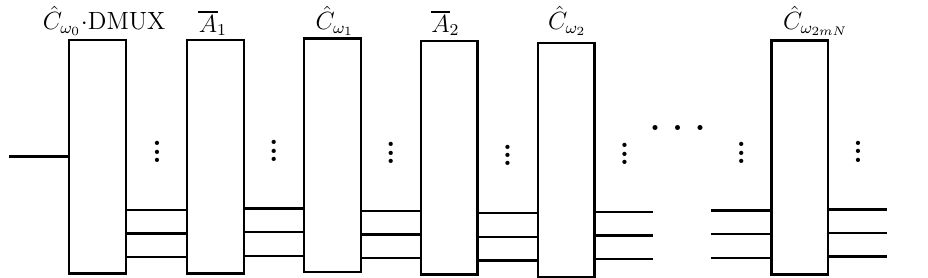}
\caption{A noise canceling protocol by repeating the noisy demultiplexer. The demultiplexer is reused for $2mN$ times ($N$ cycles of length $2m$), where the control operators $\overline{A}_j=\text{DMUX}\cdot\oper{g}_j\oper{g}^\dagger_{j-1}$ (where DMUX part is taken from the next noisy DMUX to single out the noise operator $\oper{C}_{\omega_{j+1}}$) are inserted to modify the noise generator $\oper{H}_\omega$ according to the noise cancellation principle Eq.~(\ref{eq:C-small-lambda-N}). For the first $N-1$ cycles, the control operators apply consecutively from $\oper{g}_0$ to $\oper{g}_{2m-1}$ while the last cycle they are applying in reverse order from $\oper{g}_{2m-1}$ to $\oper{g}_0$ so that the outcomes at the last stage are separated into different mode outputs.} \label{fig:protocol_generic}
\end{center}
\end{figure}

Now, from the realization above, we can summarize a working protocol for the product Eq.~(\ref{eq:C-small-lambda-N}):
\begin{enumerate}
    \item Pass the state into the noisy channel $\oper{C}_{\omega_0}\cdot$DMUX and then apply a control operator $\oper{g}_1;$ \label{begin-cycle}
    \item Pass it to the noisy channel $\oper{C}_{\omega_1}\cdot$DMUX and then apply a control operator $\oper{g}_2\oper{g}^\dagger_1;$ \label{mid-cycle}
    \item Repeat the point \ref{mid-cycle} for $2m$ times where we apply a control operator $\oper{g}_{j+1}\oper{g}^\dagger_j$ after the noisy channel $\oper{C}_{\omega_j}\cdot$DMUX; \label{j-in-cycle}
    \item Repeat the cycle \ref{begin-cycle} to \ref{j-in-cycle} for $N-1$ times;
    \item For the $N$th cycle, we begin with applying an operator $\oper{g}_{2m-1}$ to the state and pass it to the noisy channel $\oper{C}_{\omega_{2m(N-1)}}\cdot$DMUX, and then apply a control operator $\oper{g}_{2m-2}\oper{g}^\dagger_{2m-1};$\label{begin-cycle-N}
    \item Pass it to the noisy channel $\oper{C}_{\omega_{2m(N-1)+1}}\cdot$DMUX, and then apply a control operator $\oper{g}_{2m-3}\oper{g}^\dagger_{2m-2};$ \label{mid-cycle-N}
    \item Repeat the point \ref{mid-cycle-N} for $2m$ times where we apply a control operator $\oper{g}_{2m-j-1}\oper{g}^\dagger_{2m-j}$ after the noisy channel $\oper{C}_{\omega_{2m(N-1)+j}}\cdot$DMUX; the protocol will end with the noisy channel $\oper{C}_{\omega_{2mN}}\cdot$DMUX.
\end{enumerate}
Note that the control operator for the last step in the $N$th cycle is $\oper{g}_0,$ which is an identity $\iden,$ so we did not put it in the protocol. We stress here that in the last cycle, we apply the control operators in the reverse order compared to other cycles so that the final output will be in the mode separation form (rather than a single beam of all modes as the results of the realization of the operators $\oper{g}_j$ except for $\oper{g}_0.$) The overview of the protocol is equivalently depicted in Fig.~\ref{fig:protocol_generic}. In the figure, for clarity, we write an effective control operator $\overline{A}_j=\text{DMUX}\cdot\oper{g}_j\oper{g}^\dagger_{j-1}$ when we encounter the product of the noisy DMUX and the group elements $\oper{g}_j$ in the first $N-1$ cycles and $\overline{A}_j=\text{DMUX}\cdot\oper{g}_{2m-j-1}\oper{g}^\dagger_{2m-j}$ in the $N$th cycle. 
Since in terms of operators, the DMUX acts as an identity \footnote{It is not an identity operator only when one takes the output port degree of freedom into account.}, the product 
    \begin{align}
        \oper{C}_{\omega_{2mN}}&\overline{A}_{2mN-1}\cdots\oper{C}_{\omega_1}\overline{A}_1\oper{C}_{\omega_0}\nonumber\\
        &= \prod_{j=2m-1}^{0}\cv{\oper{g}^\dagger_{j}\oper{C}_{\omega_{2m(N-1)+j}}\oper{g}_{j}}\prod_{l=1}^{N-1}\Bigg[\prod_{j=0}^{2m-1}\cv{\oper{g}^\dagger_{j\oplus l}\oper{C}_{\omega_{j+l}}\oper{g}_{j\oplus l}}\Bigg]\nonumber\\ 
        &= \exp\Bigg(iN\lambda\sum_{j=0}^{2m-1}\oper{g}^\dagger_j\overline{H}\oper{g}_j\Bigg)+\mathcal{O}\cv{\lambda^2} \label{eq:C-small-lambda-N-protocol},
    \end{align}
which is simply equivalent to the product in Eq.~(\ref{eq:C-small-lambda-N}). In this sense, it is clear that the protocol we proposed above is a realization of the noise cancellation scheme for the SPADE measurement as expected.

In general, in practice, the protocol will give an approximation or partial cancellation since in principle the limit $N\rightarrow\infty$ is not feasible. However, in some special cases, such as when the noise configurations $\omega_j$ are all identical the cancellation can be achieved independent of $N$ within the regime of small $\lambda.$ 
Furthermore, {when the noise operator is additive,i.e. for any configurations $\omega$ and $\omega'$ there is a configuration $\omega''$ such that $\hat{C}_\omega\hat{C}_{\omega'} = \hat{C}_{\omega''};$ and if the commutators between noise generator $\oper{H}_\omega$ and the control operator $g_j$ are simple such that $\cvb{\oper{H}_\omega,g_j}\propto\oper{H}_\omega,$ the relation Eq.~(\ref{eq:C-small-lambda-N-protocol}) will hold without approximation and} the noise cancellation can be done perfectly. This can happen, in the example {of displacement noise} given in Sect.~\ref{sec:D-noise}, as we discuss in the following.

\subsection{Example: Subtraction of Displacement Noise in bi-SPADE Protocol}\label{sec:D-noise-mani}
\begin{figure}[!ht]
\centering
\includegraphics[width=0.5\linewidth]{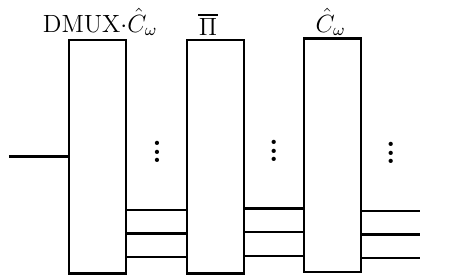}
\caption{A schematic for the example of noise cancellation protocol for bi-SPADE measurements. Here the only non-trivial control operator is a parity operator $\oper{\Pi}$ defined as $\oper{\Pi}=e^{i\pi\oper{a}^\dagger\oper{a}}$ and $\overline{\Pi}=\text{DMUX}\cdot\oper{\Pi}.$ Here the noise configurations are identical and hence the cancellation can be achieved perfectly with only two repetitions.}\label{fig:protocol_bi}
\end{figure}

Let us revisit the example in Sect.~\ref{sec:D-noise} and demonstrate that superresolution can be recovered from the displacement type disturbance with our proposed protocol. To put this in the same perspective we note that it is the case when $m=1$, and where $i\oper{H}_\omega=\overline{z}_\omega\oper{a}-z_\omega\oper{a}^\dagger$ or $\oper{C}_\omega=\oper{D}\cv{z_\omega}.$ In this sense, we can say that the noise space is $\mathcal{I}_S=\cvc{\oper{D}\cv{z}:z\in \mathbb{C}}.$ Now, we define a parity operator 
    \begin{equation}
        \oper{\Pi}=\sum_{k\in\cvb{0}_2}\ketbra{k}{k}-\sum_{k'\in\cvb{1}_2}\ketbra{k'}{k'}=e^{i\pi\oper{a}^\dagger\oper{a}} \label{eq:oper-Pi},
    \end{equation}
where $\cvb{0}_2$ ($\cvb{1}_2$) are the classes of even (odd) modes. Note that the group $\mathcal{G}_1$ is composed of an identity operator and the parity operator, $\mathcal{G}_1 = \{\iden,\oper{\Pi}\},$ i.e., $\oper{g}_0=\iden$ and $\oper{g}_1=\oper{\Pi}$ respectively. 

\subsubsection{Identical Noise Configuration}\label{sec:iden-noise-eg}
To illustrate this, we begin with the case of identical noise configuration. In practice, this may be achieved by reusing the same multiplexer for every repetition and assuming that the time duration between two consecutive repetitions is small so that the noise is frozen in the same realization. The noise cancelation mechanism in the previous subsection for this case can be simplified to the schematics in Fig.~\ref{fig:protocol_bi}. 
In this scheme, one only requires one parity operator in the middle step since in the second step the action of the operator is trivial (up to a sign) thanks to the measurement basis, i.e., $\oper{\Pi}\ket{n}=\cv{-1}^n\ket{n}.$ With the scheme depicted similar to Fig.~\ref{fig:g1_lens}, the optical implementation of the operator $\oper{\Pi}$ can be done with a lens, by placing the input and output planes at the distance $2f$ from the lens. 

First we replace this modulated measurement in Eq.~(\ref{eq:integration-compact-form}), and the measurement probability will read
    \begin{align}
        \prob_+\cv{n\vert d,\oper{C}} &=\int_{\Omega} \big\vert\bra{n}\oper{D}\cv{z_\omega}\oper{\Pi}\oper{D}\cv{ z_\omega}\ket{d}\big\vert^2d\mu\cv{\omega}\nonumber\\
             &=\int_{\Omega} \big\vert\bra{n}\oper{\Pi}\oper{D}\cv{z_\omega}\oper{\Pi}\oper{D}\cv{z_\omega}\ket{d}\big\vert^2d\mu\cv{\omega}\label{eq:P-modulated-prod}.
    \end{align}
The parity operator in Eq.~(\ref{eq:P-modulated-prod}) comes from the property $\oper{\Pi}\ket{n}=\cv{-1}^n\ket{n}.$ The product here is a special case of the product on the left-hand side of Eq.~(\ref{eq:C-small-lambda-N-protocol}), where in this case there are only two terms controlled by $\oper{g}_0=\iden$ and $\oper{g}_1=\oper{\Pi}$ respectively. 
Now, we note that the control operators $\oper{\Pi}$ and $\iden$ can then be passed to the generator of the displacements, thanks to the unitarity of the control operators. Particularly, we have
    \begin{align}
        \oper{\Pi}\oper{D}\cv{z_\omega}\oper{\Pi} &= \oper{\Pi}e^{z_\omega\oper{a}^\dagger-\overline{z}_\omega\oper{a}}\oper{\Pi}\nonumber\\
            &=\sum_{k=0}^\infty \frac{1}{k!} \oper{\Pi}\cv{z_\omega\oper{a}^\dagger-\overline{z}_\omega\oper{a}}^k\oper{\Pi}\nonumber\\
            &= \sum_{k=0}^\infty \frac{1}{k!} \cv{z_\omega\oper{\Pi}\oper{a}^\dagger\oper{\Pi}-\overline{z}_\omega\oper{\Pi}\oper{a}\oper{\Pi}}^k\nonumber\\
            &=e^{z_\omega\oper{\Pi}\oper{a}^\dagger\oper{\Pi}-\overline{z}_\omega\oper{\Pi}\oper{a}\oper{\Pi}} \label{eq:Pi-D-Pi},
    \end{align}
where we employ an identity $\oper{\Pi}^2=\iden,$ interlacing the power of the generator yielding the second last inequality. 
Let $\cvdr{\oper{a}}$ denote a $\mathcal{G}_1-$average of the annihilation operator $\cvdr{\oper{a}}:=\frac{1}{2}\big(\oper{\Pi}\oper{a}\oper{\Pi}+\oper{a}\big).$ This then leads to 
    \begin{align}
    \prob_+\cv{n\vert d,\oper{C}} &=\int_{\Omega} \Big\vert\bra{n}\exp\big[2z_\omega\cvdr{\oper{a^\dagger}}-2\overline{z}_\omega\cvdr{\oper{a}}\big]\ket{d}\Big\vert^2d\mu\cv{\omega}\label{eq:P-modulated-ave},
\end{align}
 by combining the operators' product in Eq~\ref{eq:P-modulated-prod}. This can be done since the phase is irrelevant to the absolute value, exemplifying the situation when the modified generator can be achieved without approximation.

At this point, the noise self-cancellation will automatically follow from the fact that the $\mathcal{G}_1-$average of the annihilation operator vanishes, namely
    \begin{equation}
        \cvdr{\oper{a}}=\frac{1}{2}\big(\oper{\Pi}\oper{a}\oper{\Pi}+\oper{a}\big)=\frac{1}{2}\big(-\oper{a}+\oper{a}\big)=0, \label{eq:G_1-ave-a}
    \end{equation}
and so does that of the creation operator, i.e.,  $\cvdr{\oper{a}^\dagger}=0.$ These two properties are special cases of Eq.~(\ref{eq:ap-erase}). Finally, putting this in Eq.~(\ref{eq:P-modulated-ave}), we obtain
\begin{align}
    \prob_+\cv{n\vert d,\oper{C}}&= \prob\cv{n\vert d}\label{eq:P-modulated}.
\end{align}
In short, self-cancellation of the displacement noise is achieved in the case of identical noise configuration.

For the non-static noise, where the noise configurations in the first and second steps are not identical, one can observe from Eq.~(\ref{eq:P-modulated}) that the subtraction of noise parameters $z_\omega$ that the two-step demultiplexing with the intervention of the parity operator may not be sufficient to cancel the noise in some situations. For instance, if the noise distributions at the first and second steps are uncorrelated, and each of which admits the Gaussian distribution density Eq.~(\ref{eq:noise-gaussian-profile}), the variance of the resulting probability from the manipulation scheme will be twice the variance of the individual distribution, leading to less accuracy of the estimation.

However, by the assumption that the noise in the first and second steps are generated from the same noisy demultiplexer, the increase of the variance in the situation above becomes nonphysical. 
In fact, the principle behind the cancellation mechanism relies on the strong correlation across time, hence the most correlated signal (identical noise configuration) leads to complete self-cancellation. Such correlation can be determined from the physical nature of the demultiplexer and its dynamics, so the noise configurations at two time-steps are connected by a stochastic transition given by the dynamics of the demultiplexer. 

\subsubsection{Correlated Noise Configurations}\label{sec:correlated-noise-eg}
To put the above considerations  into a quantitative picture, let us consider a joint distribution density $\mu\cv{\omega_1,\omega_2}$ whose marginals coincide with one-time noise distribution densities, i.e. {$\int_{\mathbb{C}} \mu\cv{\omega_1,\omega_2}dz_{\omega_1}=\mu\cv{\omega_2}$ and $\int_{\mathbb{C}} \mu\cv{\omega_1,\omega_2}dz_{\omega_2}=\mu\cv{\omega_1}.$} We can then rewrite Eq.~(\ref{eq:P-modulated}) with a slight modification for this general case as
\begin{align}
    \prob_+&\cv{n\vert d,\oper{C}}=\int_{\Omega}\int_{\Omega} \big\vert\bra{n}\oper{D}\cv{d-z_{\omega_2}+ z_{\omega_1}}\ket{0}\big\vert^2d\mu\cv{\omega_1,\omega_2}\label{eq:P-modulated-joint}.
\end{align}
For the ideal case of identical noise realization, we have $\mu\cv{\omega_1,\omega_2}=\delta\cv{\cvv{z_{\omega_1}-z_{\omega_1}}}\mu\cv{\omega_1}$ where $\delta$ is a Dirac delta distribution, and then Eq.~(\ref{eq:P-modulated-joint}) will reduce to Eq.~(\ref{eq:P-modulated}); whereas we would have $\mu\cv{\omega_1,\omega_2}=\mu\cv{\omega_1}\mu\cv{\omega_2}$ if the noise configurations are uncorrelated. 
Suppose that the individual distributions $\mu\cv{\omega_1}$ and $\mu\cv{\omega_2}$ are Gaussian distribution of the form Eq.~(\ref{eq:noise-gaussian-profile}). The simplest joint probability for these marginals is a bi-variate Gaussian distribution density
    \begin{align}
        \mu\cv{\omega_1,\omega_2}&= \bigg(\frac{1}{2\pi\tilde{\sigma}^2\epsilon}\bigg)^2\exp\Bigg[-\frac{\cvv{z_{\omega_1}}^2+\cvv{z_{\omega_2}}^2}{2\tilde{\sigma}^2\epsilon^2}\Bigg]\exp\Bigg[\frac{\sqrt{1-\epsilon^2}\cv{z_{\omega_1}\overline{z}_{\omega_2}+\overline{z}_{\omega_1}z_{\omega_2}}}{2\tilde{\sigma}^2\epsilon^2}\Bigg] \label{eq:example-joint},\\
        &= \bigg(\frac{1}{2\pi\tilde{\sigma}^2\epsilon}\bigg)^2\exp\Bigg[-\frac{\cvv{\xi^-_{\omega}}^2}{4\tilde{\sigma}^2\cv{1-\sqrt{1-\epsilon^2}}}\Bigg]\exp\Bigg[-\frac{\cvv{\xi^+_{\omega}}^2}{4\tilde{\sigma}^2\cv{1+\sqrt{1-\epsilon^2}}}\Bigg] \label{eq:example-joint-rotate},
    \end{align}
for $0<\epsilon<1,$ where we write $\xi^\pm_{\omega}=z_{\omega_2}\pm z_{\omega_1}.$ Here $\epsilon=1$ corresponds to the case with uncorrelated noise and $\epsilon\rightarrow 0$ approaches the identical noise configurations case (we omit the case with anti-correlation in our consideration.) Note that given two marginals the associated joint probability is not unique, and in practice, it should be constructed from the physical nature of the demultiplexer. We employ the form above, put aside the discussion on the physical nature, mainly for the sake of illustration of our scheme, and postpone a more accurate model for further studies.

Now Eq.~(\ref{eq:P-modulated-joint}) can be written as
\begin{align}
    \prob_+\cv{n\vert d,\oper{C}} &=\frac{1}{2\pi{\sigma_{\epsilon}}^2}\int_{\mathbb{C}}\big\vert\bra{n}\oper{D}\cv{d-\xi^-_{\omega}}\ket{0}\big\vert^2e^{-\frac{\cvv{\xi^-_{\omega}}^2}{2\pi{\sigma_{\epsilon}}^2}} d\xi^-_{\omega} \label{eq:prob-vary-epsilon},
\end{align}
where ${\sigma_{\epsilon}}^2=2\tilde{\sigma}^2\cv{1-\sqrt{1-\epsilon^2}}.$ Here one can see that with this type of noise correlation, the resulting probability in our modification takes the same form as Eq.~(\ref{eq:prob-Gauss-shift}) with the modification of effective noise variance $\tilde{\sigma}^2\mapsto\sigma^2_\epsilon.$ In other words, one can write
\begin{align}
    \prob_+\cv{n\vert d,\oper{C}}&=\Bigg[\frac{\partial^n}{\partial^n a}\frac{\cv{-1}^n\exp\cv{-\frac{ad^2}{1+2a\sigma_{\epsilon}^2}}}{n!\cv{1+2a\sigma_{\epsilon}^2}} \Bigg]_{a=1}  \label{eq:prob-Gauss-shift-mani}.
\end{align}
The probability above can be derived following the same derivation for Eq.~(\ref{eq:prob-Gauss-shift}). Here one can easily see that the noise correlation plays an important role in the improvement of measurement quality. Ideally, the perfect decoupling as in Eq.~(\ref{eq:P-modulated}) can be approached as $\epsilon\rightarrow 0.$ Our technique improves the quality (in the sense that the effective noise variance is reduced by our modification) up to the correlation parameter $\epsilon=\sqrt{3}/2,$ from which the original measurement is better. 

\begin{figure}[!ht]
\begin{center}
	\includegraphics[width=0.7\linewidth]{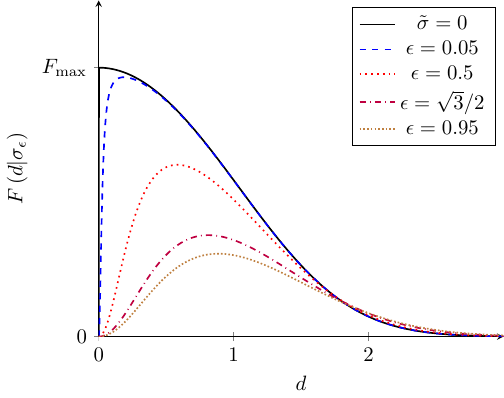}
	\caption{Fisher information in bi-SPADE protocol under Gaussian distributed displacement noise with fixed noise strength $\tilde{\sigma}=0.5,$ but various values of correlation parameter $\epsilon=0.05,0.5,\sqrt{3}/2$ and $0.95$ and hence different effective noise strengths $\sigma_\epsilon=\tilde{\sigma}\sqrt{2\cv{1-\sqrt{1-\epsilon^2}}}.$ $F_{\max}=F_0\cv{0}$ denotes the maximal Fisher information possible in the noiseless bi-SPADE procedure. At $\epsilon=\sqrt{3}/2$ the Fisher information for our modified measurement is identical to the original noisy measurement and it decreases as $\epsilon$ approaches $1.$ The resolution improves as $\epsilon$ decreases and the perfect decoupling can be reached in the limit $\epsilon\rightarrow 0.$}\label{fig:F-d-Gauss-t}
\end{center}
\end{figure}
Let us conclude the section by demonstrating explicitly how our protocol can improve the measurement sensibility for the bi-SPADE measurement. Recall the Fisher information for bi-SPADE measurement Eq.\ref{eq:F-shift-P-Gauss} with the effective noise variance
	\begin{equation}
		F\cv{d\vert \sigma_{\epsilon}} = \frac{4d^2g_{\epsilon}^3e^{-g_{\epsilon}d^2}}{1-g_{\epsilon}e^{-g_{\epsilon}d^2}} \label{eq:F-shift-P-Gauss-mani},
	\end{equation}
where $g_{\epsilon}=\frac{1}{1+\sigma_{\epsilon}^2}.$ With this modification from Fig.~\ref{fig:F-d-Gauss-t} it can be seen that as the parameter $\epsilon$ decreases, the resolution improves, and within the limit $\epsilon\rightarrow 0$ the Fisher information becomes closer to the noiseless case as claimed. The limit for the distribution is simply the expression Eq.~(\ref{eq:P-modulated}) where the noise configurations at two time steps are treated identically and the perfect decoupling can be achieved. For non-zero $\epsilon,$ although the singularity still remains, one can observe that the protocol substantially improves the sensitivity, suggesting that one can gain higher measurement resolution in the sub-Rayleigh's regime.

%%%%%%%%%%%%%%%%%%%%%%%%%%%%%%%%%%%%%%%%%%%%%%%%%%%%%%
%%%% CONCLUSIONS	
%%%%%%%%%%%%%%%%%%%%%%%%%%%%%%%%%%%%%%%%%%%%%%%%%%%%%%
\section{Conclusions}\label{sec:conclusion}
We have analyzed noisy SPADE measurements for resolving the transverse spatial degree of freedom of a photon beam. We have considered a noise model given as an extra layer of a random unitary channel whose generator is described by a polynomial in the creation and annihilation operators. We have shown that a group of rotations on the phase space can be considered as control operators, in which their cyclic interventions interlacing the repetitions of the mode separators or demultiplexers will modify the effective noise generators in such a way that the noise is canceled in the limit of large number of repetitions and small noise strength. This mechanism can be implemented within the spirit of the standard dynamical decoupling, where the product of noise channels intervened by the rotations can be interpreted as an analog of a noisy dynamical map perturbed by a set of control operators for which the noise contribution is absent in the effective Hamiltonian. 

We have briefly discussed the operational description of such control operators in terms of mode modulators and ray optics manipulations, and then we propose a noise-canceling protocol for SPADE measurements. For generic cases, in the limit of a large number of repetition and small noise strength, the perfect decoupling can be achieved as a limit of the protocol. For a special case when the commutators of the noise operators and the control operators are constant, and the noise moreover does not vary in time steps, the protocol can be simplified and the limit procedure can be omitted. We have demonstrated such a situation with an example of displacement noise and have shown that the perfect decoupling of the noise can be achieved for bi-SPADE measurement when the noise configurations in the two repetitions of the demultiplexer are identical. For less correlated noise we have found that our protocol provides an improved resolution in the  sub-Rayleigh regime. 

According to our results, one can apply the noise decoupling mechanism principle from the manipulation of noisy time evolution (frequency-time degrees of freedom) to the control of noise with other CVs, i.e. position and momentum degrees of freedom. It is interesting that other similar topics related to the dynamical decoupling protocol can be translated into this setup as well. For instance, noise spectroscopy and filter base recovery can be also achieved for position-momentum variables. This is useful for situations when the noise contribution cannot be erased completely, so by reconstructing a noise contribution from a known state subject to modulated noise channels given by our protocol, one can use it to clarify the measured state by demodulation with such a reconstructed filter. This example, as well as similar ones, are interesting topics to consider for further studies concerning noise suppression in metrology with continuous variables.

\backmatter

%\bmhead{Supplementary information}
%
%If your article has accompanying supplementary file/s please state so here. 
%
%Authors reporting data from electrophoretic gels and blots should supply the full unprocessed scans for key as part of their Supplementary information. This may be requested by the editorial team/s if it is missing.
%
%Please refer to Journal-level guidance for any specific requirements.

\bmhead{Acknowledgments}
Comments and suggestions from  B. Tonekaboni, M. Markiewicz, T. Linowski, K. Schlichtholz and T. Chalermpusitarak on the development of this manuscript are gratefully appreciated. We thank Mattia Walschaers for pointing out Ref. \cite{Cleme}. We acknowledge support from the Foundation for Polish Science (IRAP project, ICTQT, contract no. 2018/MAB/5, co-financed by EU within Smart Growth Operational Programme). This work is partially carried out under IRA Programme, project no. FENG.02.01-IP.05-0006/23, financed by the FENG program 2021-2027, Priority FENG.02, Measure FENG.02.01., with the support of the FNP.

%\bibliography{reference.bib}

\begin{thebibliography}{10}
\providecommand{\url}[1]{{#1}}
\providecommand{\urlprefix}{URL }
\providecommand{\doi}[1]{\url{https://doi.org/#1}}
\bibcommenthead

\bibitem{Tsang2016}
M.~Tsang, R.~Nair, X.M. Lu, Quantum theory of superresolution for two incoherent optical point sources.
\newblock Phys.Rev.X \textbf{6}, 031,033 (2016).
\newblock \doi{10.1103/PhysRevX.6.031033}.
\newblock \urlprefix\url{https://link.aps.org/doi/10.1103/PhysRevX.6.031033}

\bibitem{Nair2016}
R.~Nair, M.~Tsang, Far-field superresolution of thermal electromagnetic sources at the quantum limit.
\newblock Phys. Rev. Lett. \textbf{117}, 190,801 (2016).
\newblock \doi{10.1103/PhysRevLett.117.190801}.
\newblock \urlprefix\url{https://link.aps.org/doi/10.1103/PhysRevLett.117.190801}

\bibitem{Lupo2016}
C.~Lupo, S.~Pirandola, Ultimate precision bound of quantum and subwavelength imaging.
\newblock Phys. Rev. Lett. \textbf{117}, 190,802 (2016).
\newblock \doi{10.1103/PhysRevLett.117.190802}.
\newblock \urlprefix\url{https://link.aps.org/doi/10.1103/PhysRevLett.117.190802}

\bibitem{Ang2017}
S.Z. Ang, R.~Nair, M.~Tsang, Quantum limit for two-dimensional resolution of two incoherent optical point sources.
\newblock Phys. Rev. A \textbf{95}, 063,847 (2017).
\newblock \doi{10.1103/PhysRevA.95.063847}.
\newblock \urlprefix\url{https://link.aps.org/doi/10.1103/PhysRevA.95.063847}

\bibitem{Yu2018}
Z.~Yu, S.~Prasad, Quantum limited superresolution of an incoherent source pair in three dimensions.
\newblock Phys. Rev. Lett. \textbf{121}, 180,504 (2018).
\newblock \doi{10.1103/PhysRevLett.121.180504}.
\newblock \urlprefix\url{https://link.aps.org/doi/10.1103/PhysRevLett.121.180504}

\bibitem{Yang2019}
J.~Yang, S.~Pang, Y.~Zhou, A.N. Jordan, Optimal measurements for quantum multiparameter estimation with general states.
\newblock Phys. Rev. A \textbf{100}, 032,104 (2019).
\newblock \doi{10.1103/PhysRevA.100.032104}.
\newblock \urlprefix\url{https://link.aps.org/doi/10.1103/PhysRevA.100.032104}

\bibitem{Peng2021}
L.~Peng, X.M. Lu, Generalization of rayleigh's criterion on parameter estimation with incoherent sources.
\newblock Phys. Rev. A \textbf{103}, 042,601 (2021).
\newblock \doi{10.1103/PhysRevA.103.042601}.
\newblock \urlprefix\url{https://link.aps.org/doi/10.1103/PhysRevA.103.042601}

\bibitem{Matlin2022}
E.F. Matlin, L.J. Zipp, Imaging arbitrary incoherent source distributions with near quantum-limited resolution.
\newblock Scientific Reports \textbf{12}(1) (2022).
\newblock \doi{10.1038/s41598-022-06644-3}.
\newblock \urlprefix\url{https://doi.org/10.1038/s41598-022-06644-3}

\bibitem{Paur2016}
M.~Pa\'{u}r, B.~Stoklasa, Z.~Hradil, L.L. S\'{a}nchez-Soto, J.~Rehacek, Achieving the ultimate optical resolution.
\newblock Optica \textbf{3}(10), 1144--1147 (2016).
\newblock \doi{10.1364/OPTICA.3.001144}.
\newblock \urlprefix\url{http://opg.optica.org/optica/abstract.cfm?URI=optica-3-10-1144}

\bibitem{Tham2017}
W.K. Tham, H.~Ferretti, A.M. Steinberg, Beating rayleigh's curse by imaging using phase information.
\newblock Phys. Rev. Lett. \textbf{118}, 070,801 (2017).
\newblock \doi{10.1103/PhysRevLett.118.070801}.
\newblock \urlprefix\url{https://link.aps.org/doi/10.1103/PhysRevLett.118.070801}

\bibitem{Zhou2019}
Y.~Zhou, J.~Yang, J.D. Hassett, S.M.H. Rafsanjani, M.~Mirhosseini, A.N. Vamivakas, A.N. Jordan, Z.~Shi, R.W. Boyd, Quantum-limited estimation of the axial separation of two incoherent point sources.
\newblock Optica \textbf{6}(5), 534--541 (2019).
\newblock \doi{10.1364/OPTICA.6.000534}.
\newblock \urlprefix\url{http://opg.optica.org/optica/abstract.cfm?URI=optica-6-5-534}

\bibitem{Zhang2020}
H.~Zhang, S.~Kumar, Y.P. Huang, Super-resolution optical classifier with high photon efficiency.
\newblock Opt. Lett. \textbf{45}(18), 4968--4971 (2020).
\newblock \doi{10.1364/OL.401614}.
\newblock \urlprefix\url{http://opg.optica.org/ol/abstract.cfm?URI=ol-45-18-4968}

\bibitem{Joshi2022}
C.~Joshi, B.M. Sparkes, A.~Farsi, T.~Gerrits, V.~Verma, S.~Ramelow, S.W. Nam, A.L. Gaeta, Picosecond-resolution single-photon time lens for temporal mode quantum processing.
\newblock Optica \textbf{9}(4), 364--373 (2022).
\newblock \doi{10.1364/OPTICA.439827}.
\newblock \urlprefix\url{http://opg.optica.org/optica/abstract.cfm?URI=optica-9-4-364}

\bibitem{Cleme}
C.~Rouvi\`{e}re, D.~Barral, A.~Grateau, I.~Karuseichyk, G.~Sorelli, M.~Walschaers, N.~Treps, Ultra-sensitive separation estimation of optical sources.
\newblock Optica \textbf{11}(2), 166--170 (2024).
\newblock \doi{10.1364/OPTICA.500039}

\bibitem{Gessner2020}
M.~Gessner, C.~Fabre, N.~Treps, Superresolution limits from measurement crosstalk.
\newblock Phys.Rev.Lett. \textbf{125}, 100,501 (2020).
\newblock \doi{10.1103/PhysRevLett.125.100501}.
\newblock \urlprefix\url{https://link.aps.org/doi/10.1103/PhysRevLett.125.100501}

\bibitem{Len2020}
Y.L. Len, C.~Datta, M.~Parniak, K.~Banaszek, Resolution limits of spatial mode demultiplexing with noisy detection.
\newblock International Journal of Quantum Information \textbf{18}(01), 1941,015 (2020).
\newblock \doi{10.1142/s0219749919410156}.
\newblock \urlprefix\url{https://doi.org/10.1142/s0219749919410156}

\bibitem{Oh2021}
C.~Oh, S.~Zhou, Y.~Wong, L.~Jiang, Quantum limits of superresolution in a noisy environment.
\newblock Phys. Rev. Lett. \textbf{126}, 120,502 (2021).
\newblock \doi{10.1103/PhysRevLett.126.120502}.
\newblock \urlprefix\url{https://link.aps.org/doi/10.1103/PhysRevLett.126.120502}

\bibitem{Viola1998}
L.~Viola, S.~Lloyd, Dynamical suppression of decoherence in two-state quantum systems.
\newblock Phys. Rev. A \textbf{58}, 2733--2744 (1998).
\newblock \doi{10.1103/PhysRevA.58.2733}.
\newblock \urlprefix\url{https://link.aps.org/doi/10.1103/PhysRevA.58.2733}

\bibitem{Viola1999a}
L.~Viola, E.~Knill, S.~Lloyd, Dynamical decoupling of open quantum systems.
\newblock Phys. Rev. Lett. \textbf{82}, 2417--2421 (1999).
\newblock \doi{10.1103/PhysRevLett.82.2417}.
\newblock \urlprefix\url{https://link.aps.org/doi/10.1103/PhysRevLett.82.2417}

\bibitem{Viola1999b}
L.~Viola, S.~Lloyd, E.~Knill, Universal control of decoupled quantum systems.
\newblock Phys. Rev. Lett. \textbf{83}, 4888--4891 (1999).
\newblock \doi{10.1103/PhysRevLett.83.4888}.
\newblock \urlprefix\url{https://link.aps.org/doi/10.1103/PhysRevLett.83.4888}

\bibitem{Sakuldee2024}
F.~Sakuldee, B.~Tonekaboni, Noise decoupling for state transfer in continuous-variable systems.
\newblock Phys. Rev. A \textbf{109}, 032,404 (2024).
\newblock \doi{10.1103/PhysRevA.109.032404}.
\newblock \urlprefix\url{https://link.aps.org/doi/10.1103/PhysRevA.109.032404}

\bibitem{Rossi2018}
R.J. Rossi, \emph{Mathematical Statistics. An Introduction to Likelihood based Inference} (Wiley, 2018)

\bibitem{crameer1946}
H.~Crame{\'e}r, \emph{Mathematical methods of statistics} (Princeton University Press, Princeton, 1946)

\bibitem{Rao1992}
C.R. Rao, in \emph{Springer Series in Statistics} (Springer New York, 1992), pp. 235--247.
\newblock \doi{10.1007/978-1-4612-0919-5_16}.
\newblock \urlprefix\url{https://doi.org/10.1007/978-1-4612-0919-5_16}

\bibitem{Linowski2023}
T.~Linowski, K.~Schlichtholz, G.~Sorelli, M.~Gessner, M.~Walschaers, N.~Treps, Łukasz Rudnicki, Application range of crosstalk-affected spatial demultiplexing for resolving separations between unbalanced sources.
\newblock New Journal of Physics \textbf{25}(10), 103,050 (2023).
\newblock \doi{10.1088/1367-2630/ad0173}.
\newblock \urlprefix\url{https://dx.doi.org/10.1088/1367-2630/ad0173}

\bibitem{deAlmeida2021}
J.O. de~Almeida, J.~Ko\l{}ody\ifmmode~\acute{n}\else \'{n}\fi{}ski, C.~Hirche, M.~Lewenstein, M.~Skotiniotis, Discrimination and estimation of incoherent sources under misalignment.
\newblock Phys. Rev. A \textbf{103}, 022,406 (2021).
\newblock \doi{10.1103/PhysRevA.103.022406}.
\newblock \urlprefix\url{https://link.aps.org/doi/10.1103/PhysRevA.103.022406}

\bibitem{Schlichtholz2024}
K.~Schlichtholz, T.~Linowski, M.~Walschaers, N.~Treps, {\L}.~Rudnicki, G.~Sorelli, Practical tests for sub-rayleigh source discriminations with imperfect demultiplexers.
\newblock Optica Quantum \textbf{2}(1), 29--34 (2024).
\newblock \doi{10.1364/OPTICAQ.502459}.
\newblock \urlprefix\url{https://opg.optica.org/opticaq/abstract.cfm?URI=opticaq-2-1-29}

\bibitem{Stoler1981}
D.~Stoler, Operator methods in physical optics.
\newblock J. Opt. Soc. Am. \textbf{71}(3), 334--341 (1981).
\newblock \doi{10.1364/JOSA.71.000334}.
\newblock \urlprefix\url{https://opg.optica.org/abstract.cfm?URI=josa-71-3-334}

\bibitem{Ozaktas1995}
H.M. Ozaktas, D.~Mendlovic, Fractional fourier optics.
\newblock J. Opt. Soc. Am. A \textbf{12}(4), 743--751 (1995).
\newblock \doi{10.1364/JOSAA.12.000743}.
\newblock \urlprefix\url{https://opg.optica.org/josaa/abstract.cfm?URI=josaa-12-4-743}

\bibitem{Jagoszewski1998}
E.~Jagoszewski, {Fractional fourier transform in optical setups}.
\newblock Optica Applicata \textbf{28}(3), 236--237 (1998)

\end{thebibliography}

\end{document}